\documentclass[twocolumn]{aastex62}
\usepackage{listings}
\usepackage{graphicx}
\usepackage{dcolumn}
\usepackage{bm}
\usepackage{float,color}
\usepackage{xcolor}
\usepackage[normalem]{ulem}
\usepackage{tabularx}
\usepackage{mathtools} 
\usepackage{multirow}
\usepackage{scrextend}
\usepackage{hyperref}

\begin{document}
\title{ 
Remnant Black Hole Kicks and Implications for Hierarchical Mergers
}

\correspondingauthor{Parthapratim Mahapatra}
\email{ppmp75@cmi.ac.in}

\author[0000-0002-5490-2558]{Parthapratim Mahapatra}
\affiliation{Chennai Mathematical Institute, Siruseri, 603103, India}
\author[0000-0002-5441-9013]{Anuradha Gupta}
\affiliation{Department of Physics and Astronomy, The University of Mississippi, Oxford MS 38677, USA}
\author[0000-0001-8270-9512]{Marc Favata}
\affiliation{Department of Physics \& Astronomy, Montclair State University, 1 Normal Avenue, Montclair, NJ 07043, USA}
\author[0000-0002-6960-8538]{K. G. Arun}
\affiliation{Chennai Mathematical Institute, Siruseri, 603103, India}
\author[0000-0003-3845-7586]{B. S. Sathyaprakash}
\affiliation{Institute for Gravitation and the Cosmos, Department of Physics, Penn State University, University Park, PA 16802, USA}
\affiliation{Department of Astronomy and Astrophysics, Penn State University, University Park, PA 16802, USA}
\affiliation{School of Physics and Astronomy, Cardiff University, Cardiff, CF24 3AA, United Kingdom}

\begin{abstract}
When binary black holes merge in dense star clusters, their remnants can pair up with other black holes in the cluster, forming heavier and heavier black holes in a process called \emph{hierarchical merger.} The most important condition for hierarchical merger to occur is that remnants formed by mergers are retained by the host star cluster. Using the publicly available gravitational-wave event database, we infer the magnitudes of kick velocities  imparted to the remnant black holes due to anisotropic emission of gravitational waves and use that to quantify the retention probability of each event as a function of the escape speed of the star cluster. Among the second gravitational-wave transient catalog (GWTC-2) events, GW190814 provides the tightest constraint on the kick magnitude with ${\rm V_{kick}}=74_{-7}^{+10}$ km/s at the 90\% credible level. We find that star clusters with escape speeds of 200 km/s can retain about 50\% of the events in the GWTC-2. 
Using the escape speed distributions of nuclear star clusters and globular clusters, we find that $\sim 17$ (2) remnants of GWTC-2 may be retained by the host star cluster if all GWTC-2 events occurred in nuclear (globular) clusters.   Our study demonstrates the importance of folding in kick velocity inferences in future studies of hierarchical mergers.
\end{abstract}

\section{Introduction} \label{sec:intro}
 Several of the stellar-mass compact binary mergers observed to date could have occurred in gravitationally bound environments, such as globular clusters (GCs) or nuclear star clusters (NSCs) \citep{PortegiesZwart:1999nm,Miller:2008yw,Downing:2009ag,Rodriguez:2015oxa,Rodriguez:2016kxx,Rodriguez:2018pss,Petrovich:2017otm,Sedda2018,Samsing:2017xmd,Fragione:2018vty,Zevin:2018kzq,Antonini:2019ulv}. If the star cluster is dense enough, the remnant black hole formed by a merger can subsequently pair with another black hole (BH) via dynamical interactions, merging again under gravitational-wave radiation reaction \citep{Doctor:2019ruh,Antonini:2018auk,Fragione:2020nib,Mapelli:2020xeq}. This can repeat, producing increasingly more massive black holes at each step---a process called  \emph{hierarchical mergers} \citep{Fragione:2017blf,Fragione:2018lvy}. 
 Evidence of significant numbers of black holes formed via hierarchical mergers would strongly suggest that compact object binaries form dynamically in dense star clusters. This process can also lead to the formation of intermediate mass black holes \citep{Miller:2001ez}, which could in turn seed the formation of supermassive black holes~\citep{Bellovary:2019nib}.
 
 The presence of \emph{heavy} BHs (those with masses $>50 M_{\odot}$) in the second LIGO/Virgo Collaboration (LVC) Gravitational-Wave Transient Catalog (GWTC-2)~\citep{GWTC2} has led to detailed investigation into hierarchical mergers \citep{Gerosa:2020bjb, Kimball:2020opk, Tiwari:2020otp}. Recent analyses~\citep{Kimball:2020qyd} have found  evidence  of hierarchical mergers in six  GWTC-2 events. This inference relies on assessing the consistency of the mass and spin parameter distributions with those expected from hierarchical mergers~\citep{Fishbach:2017dwv,Gerosa:2017kvu}.

 It is well known that the anisotropic emission of gravitational waves (GWs) during the end-stages of binary coalescence carries away linear momentum and produces a recoil or ``kick'' of the merger remnant  \citep{Fitchett83,Favata:2004wz}. This plays a decisive role in the hierarchical merger process, which cannot occur if a merger remnant is ejected from its dense stellar environment. Black hole kicks can reach hundreds to thousands of kilometers per second \citep{Campanelli:2007cga,Gonzalez:2006,Tichy:2007hk,Lousto:2011kp,Lousto:2019lyf}. In comparison, globular clusters have escape speeds $\sim 2$--$180$ km/s, and nuclear star clusters have larger escape speeds $\sim 10$--$1000$ km/s~\citep{Antonini:2016gqe}.  For the hierarchical merger process to operate, the BH kick should not exceed the cluster escape speed \citep{Merritt04}. 
 
 A direct detection of the GW kicks is very challenging as it would entail measuring either the very small Doppler shift in the frequencies of the quasi-normal modes of the remnant BHs~\citep{Favata:2008ti,Gerosa:2016vip}, the rich structure in the higher-order modes of gravitational waves~\citep{CalderonBustillo:2018zuq}, or tracking certain post-merger features in the gravitational waveforms~\citep{CalderonBustillo:2019wwe}. However, numerical relativity (NR) provides useful fitting formulas~\citep{Campanelli:2007ew}, expressing the remnant kick as a function of the mass ratio and spin configuration of the binary components. This permits inference of GW kicks for a binary black hole (BBH) merger, provided there are reasonably precise measurements of the mass and spin parameters~\citep{Varma:2020nbm}.
 
 Recently, \citet{Fragione:2020miv} computed the kick velocity distributions of GWTC-2 events as a function of the spins of the binary components and studied the probability of retaining these mergers as a function of spin magnitudes. They found that only nuclear star clusters with escape speeds higher than 100 km/s could retain the GWTC-2 merger remnants even for dimensionless spins as small as $0.1$.  
 Note that binaries containing higher-spinning BHs  typically experience larger kicks as compared to binaries with low-spin BHs, and can thus be more easily ejected from clusters.
 
 \citet{Doctor:2021qfn} mapped the properties of GWTC-2 BH remnants to the statistical properties of remnant BHs in our universe. They obtained the distribution of kick velocities of GWTC-2 events using NR surrogate waveform models \citep{Varma:2018aht,Varma:2019csw} 
 and found that globular clusters and nuclear clusters can retain $\sim 4\%$ and $45\%$ 
 of the remnants, respectively. They assumed the escape speeds of  all globular clusters to be $50$ km/s and that of all nuclear clusters to be $250$ km/s.
 
Our primary goal is to compute the posterior distributions of the GWTC-2 remnant kicks using NR fitting formulas and deduce the probability that remnant BHs are retained by their host star clusters. This, in turn, will help us determine if GWTC-2 remnants can facilitate hierarchical mergers.  

Specifically, we develop a  framework to  compute the probability that a BBH merger is retained by its  environment (assuming the merger takes place inside a star cluster and that dynamical $N$-body interactions do not play a significant role in ejecting heavy BHs). This is applied to all BBHs in the GWTC-2 catalog, obtaining their individual retention probabilities as a function of the cluster escape speed. We also discuss what the kick velocities of the GWTC-2 population imply for different types of star clusters and their efficiency to retain BH remnants. 

We find  that BH kicks will not be a major obstacle for hierarchical mergers if  star clusters with escape speeds $\gtrsim$ 200 km/s are abundant in the universe and BBHs with masses characteristic of those in GWTC-2 primarily form in such clusters. 
We also find that among the six GWTC-2 events identified by \cite{Kimball:2020qyd} as systems showing evidence for hierarchical mergers, all  except GW190517\_055101 may be retained by their host clusters with a probability $\sim 50\%$, paving the path for participating in further binary formation, provided they merged in clusters with escape speeds larger than $\sim 700$ km/s.  With our present knowledge of the escape speed distribution of star clusters, $\sim 17\, (2)$ of the GWTC-2 remnants may be retained by nuclear (globular) clusters. \emph{We stress the importance of using kick velocity measurements as an ingredient in future studies of hierarchical black hole formation.} 

\section{Inferring retention probability from kick posterior distribution}\label{sec:method}
\begin{figure*}[ht!]
    \centering
    \includegraphics[width=0.4955\textwidth]{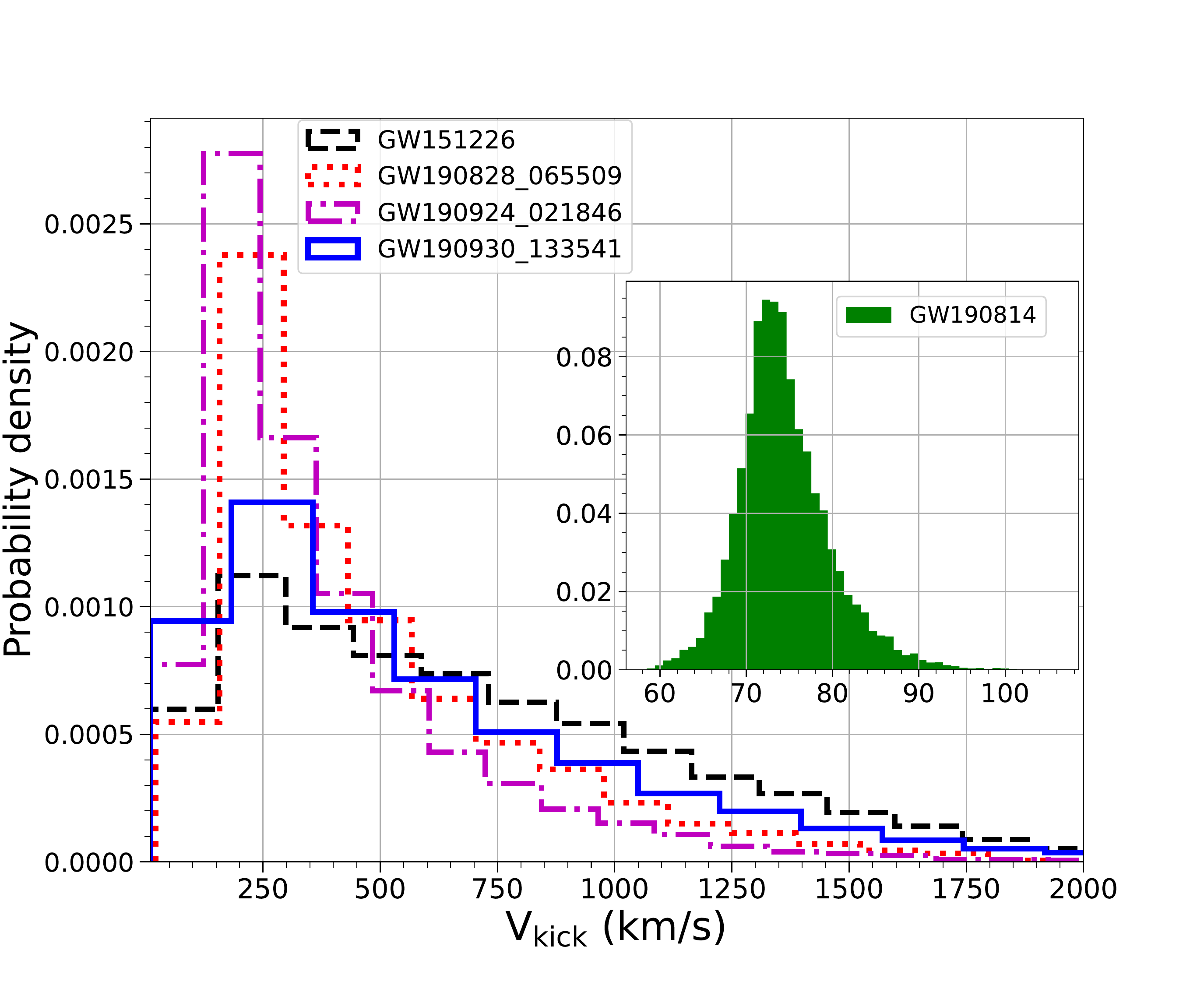}
    \includegraphics[width=0.4955\textwidth]{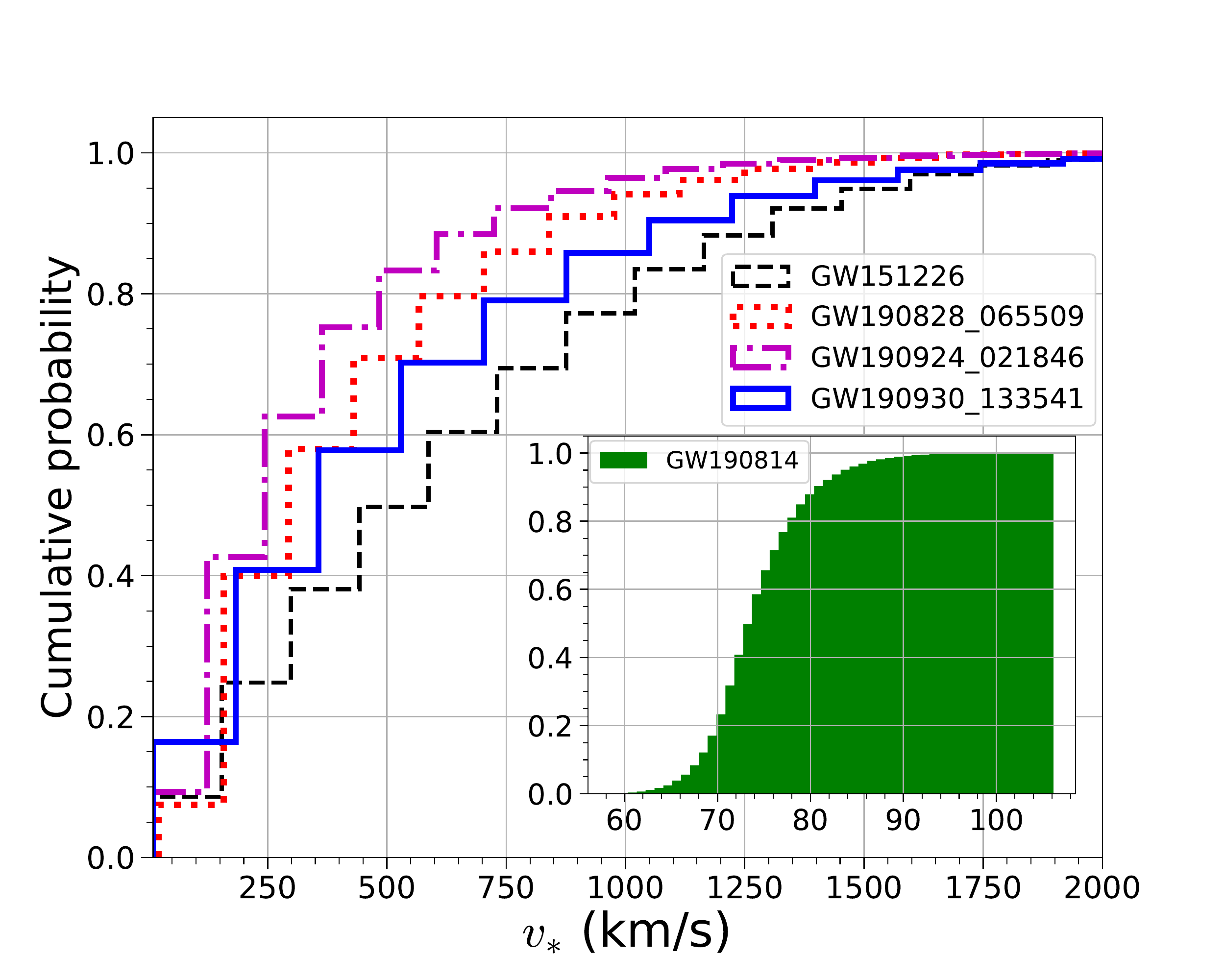}
    \caption{The probability density function (PDF) $p({\rm V_{ kick}})$ (left), and the cumulative distribution function (CDF) $F( v_{\ast})$ (right) of the kick magnitude for representative GW events in GWTC-2.  The PDF is constructed by combining the parameter estimation posterior samples for the binary parameters with the fitting formula for the kick. The CDF $F( v_{\ast})$ 
    is computed by integrating $p({\rm V_{ kick}})$ from $0$ to $v_{\ast}$. 
    Shown in the insets are the PDF and the CDF of GW190814, which has the most precisely inferred kick magnitude and the lowest kick of all GWTC-2 events.
}
    \label{fig:GW190814kickCDF}
\end{figure*}
We first discuss how we infer the kick velocity probability distribution for a given GWTC-2 event. We then compute the retention probability of each BBH event as a function of the cluster escape speed $\rm {V_{esc}}$.  
\subsection{Kick inference}
The kick imparted by the GW linear momentum loss is described by fitting formulas based on NR simulations of BBHs~\citep{Gonzalez:2006, Campanelli:2007ew, Gonzalez:2007hi, Lousto:2007, Lousto:2012su, Lousto:2013}. We apply the kick magnitude of \citet{Campanelli:2007ew}, summarized in Appendix \ref{sec:kick-appen}.

The kick velocity depends on the following parameters: the mass ratio $q$ ($\leq1$); the dimensionless spin magnitudes $\chi_{\rm i}$; the angles between the spin vectors and the total orbital angular momentum $\theta_{\rm i}$; the difference between the azimuthal angles of the two spin vectors, $\phi_{12}$; and an additional parameter $\Theta$, which is the angle between ${\bm \Delta} \times \hat{{\textbf{L}}}$ and a fiducial infall direction of the two BHs at merger (see Appendix \ref{sec:kick-appen} for the definition of  ${\bm \Delta}$ and  $\hat{{\textbf L}}$). Here ${\rm i=1, 2}$ labels the primary or secondary BH. Currently the posterior samples of $\theta_{\rm i}$ for GWTC-2 events are not available, but the posterior samples of $\theta_{LS_{\rm i}}$ (the angles between the spin vectors and the direction of the Newtonian orbital angular momentum ${\hat{\textbf{L}}}_{N}$) are accessible. Here we use the posterior samples for $\theta_{LS_{\rm i}}$, assuming $\hat{\textbf{L}}\approx {\hat{\textbf{L}}}_{N}$ and $\theta_{\rm i} \approx \theta_{LS_{\rm i}}$. Deviations in the directions of $\hat{\textbf{L}}$ and ${\hat{\textbf{L}}}_{N}$ only enter as relative 1.5 post-Newtonian-order corrections (i.e., corrections cubic in the relative orbital speed) and are proportional to the components of the spin perpendicular to ${\hat{\textbf{L}}}_{N}$ \citep{K95}; since there is little evidence for large spins or precession in GWTC-2, $\hat{\textbf{L}}\approx {\hat{\textbf{L}}}_{N}$ is a reasonable approximation.

In addition to feeding samples of $q$, $\chi_{\rm i}$, $\theta_{LS_{\rm i}}$, and $ \phi_{12}$, into the kick fitting formula, we also need to know the parameter $\Theta$. As we cannot directly infer $\Theta$ from the posterior samples, we assume it is uniformly distributed between [0, 2$\pi$] \citep{Gerosa:2014gja,Gerosa:2016sys} when computing the posteriors of the kick magnitude via Equation~(\ref{eq:kick}).

Restricting to the 47 BBHs in GWTC-2 \citep{GWTC2}, we construct probability density functions (PDFs) for the kick magnitude, using as input the posterior PDFs for the mass ratio and spin parameters of each BBH event. (See Appendix \ref{sec:kick-appen} for additional details on constructing the kick magnitude PDFs.)

Figure \ref{fig:GW190814kickCDF} shows some typical examples of the resulting kick PDFs $p({\rm V_{kick}})$ for a few selected GW events. GW190814 provides the best inference of the kick and has the lowest inferred kick. The former is due to the stringent constraint on the spin of its primary ($\chi_1\leq 0.007$), and the latter is due to the small spin magnitude.

As the spins for several GW events are poorly constrained, our analysis  risks returning posteriors that are prior dominated.  We use the Jensen-Shannon (JS) divergence~\citep{JSdiv} to quantify how informative are the posteriors compared to the priors, considering only those events that cross a JS divergence threshold of $0.007$. This leaves 42 events for further analysis. See Appendix~\ref{sec:JSdiv} for  further discussion. 

Given a  posterior PDF $p(\rm{V_{kick}})$ for the kick magnitude of each event, the corresponding cumulative distribution function (CDF) $F(v_{\ast}) = \int_{0}^{v_{\ast}} p({\rm V_{kick}})\, d {\rm V_{ kick}}$ quantifies the probability that ${\rm V_{kick}} \leq v_{\ast}$ and therefore gives the retention probability $F({\rm V_{esc}})$ given a cluster with an escape velocity ${\rm V_{esc}} = v_{\ast}$. Example CDFs are shown in  the right panel of Figure~\ref{fig:GW190814kickCDF}.
 
\subsection{Retention probability of GWTC-2 events}
Operating under the assumption that all reported GWTC-2 BBHs occur in the dense core of star clusters with unknown escape speeds, the retention probability ${\rm P}_{{\rm ret},j}(\rm{V_{esc}})$ of the $j^{\rm th}$ event in a cluster with escape speed ${\rm V_{esc}}$ is simply found by replacing $v_{\ast} \rightarrow {\rm V_{esc}}$ in the CDF; i.e.,  ${\rm P}_{{\rm ret},j}({\rm V_{esc}}) = F(\rm{V_{esc}})$.
For example, because the CDF for the GW190814 kick distribution indicates that there is $\approx 50\%$ probability that ${\rm V_{kick}}<74$ km/s (Figure~\ref{fig:GW190814kickCDF} insets), this implies that a cluster with ${\rm V_{esc}} = 74$ km/s would retain this binary with $\approx 50\%$ probability.

\begin{figure}[t]
    \centering
    \includegraphics[scale=0.45]{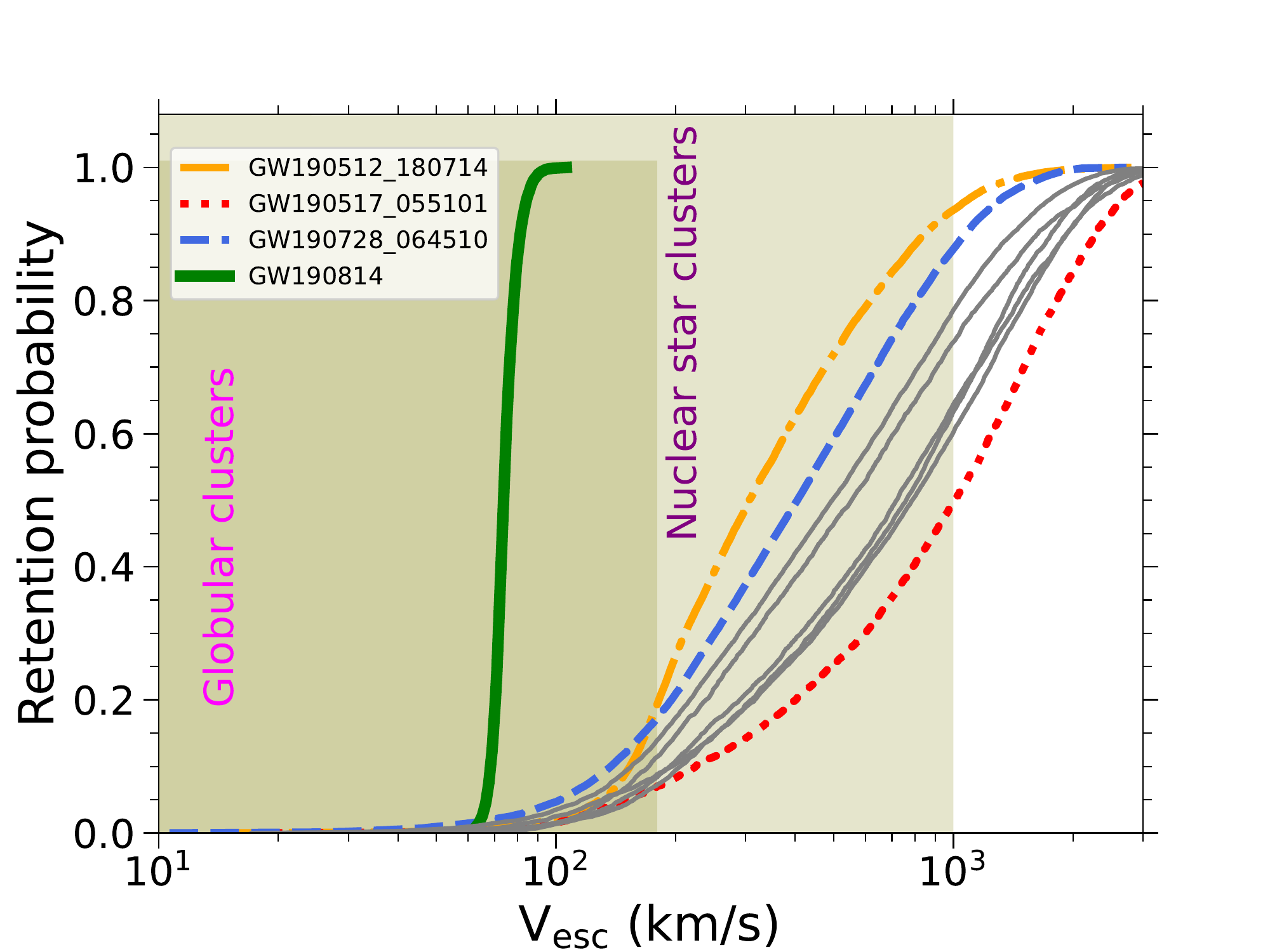}
    \caption{\label{fig:Retention0.5pc1}Retention probability of a few representative  GWTC-2 events as a function of the cluster escape speed (thick curves, listed in the legend). The gray curves correspond to five of the six events reported by \cite{Kimball:2020qyd} to be of hierarchical origin (the sixth one is the red dotted curve; see Section~\ref{sec:sixevents} for discussion). The retention probability is computed directly from the kick CDF (Figure~\ref{fig:GW190814kickCDF}) as discussed in the text.  The shaded regions show the range of escape speeds for globular clusters and nuclear star clusters \citep{Antonini:2016gqe}. For example, a nuclear star cluster with escape speed $\sim 500$ km/s can retain GW190512$\_$180714-like systems with probability $\sim 0.70$, while a globular cluster with escape speed $\sim 74$ km/s can retain a GW190814-like system with probability $\sim 0.50$. Higher retention probability increases the likelihood that merger remnants can participate in subsequent BBH mergers within the cluster. 
}
\end{figure}
We apply this observation to the  42 BBHs in GWTC-2 with informative kick posteriors, mapping the kick magnitude CDFs to their retention probabilities. This result is shown in Figure~\ref{fig:Retention0.5pc1} for  a selected sample of the  42 GWTC-2 events. (The four events highlighted in thick colored lines are representative examples of the range of kick velocity distributions; the thin gray curves are discussed below.) These retention probabilities help to quantify, as a function of the cluster escape speed, the ability of these events to take part in future BBH mergers.

It is evident that GW190814 has the highest retention probability among the GWTC-2 events with informative kick posteriors. It would almost surely be retained by a cluster with an escape speed as low as 80 km/s. The lowest retention probability is for GW190517$\_$055101, which would be retained at 50\% probability by clusters with an escape speed of $\sim 1000$ km/s.  

As these conclusions depend only weakly on the details of the clusters, our method provides a powerful diagnostic of the ability of star clusters to retain the  BBH merger remnants. Retaining these remnants is an important prerequisite if they are to participate in subsequent BBH mergers.

\subsection{\label{sec:sixevents}Retention of likely hierarchical merger events}
We now assess the retention probabilities of the six events reported in \cite{Kimball:2020qyd} and discussed earlier. Using the method of the previous section, we compute the kick PDFs for these six events and determine their retention probabilities ${\rm P}_{{\rm ret},j}({\rm V_{esc}})$ as a function of cluster escape speed. The result is shown in Figure~\ref{fig:Retention0.5pc1}, where the gray lines represent five out of the six events that \cite{Kimball:2020qyd} indicate as likely to be hierarchical in origin. The sixth event (GW190517\_055101) is shown in dotted red and has the lowest retention probability ($30\%$ for a cluster with escape speed $600 {\rm \,km/s}$).
The highest retention probability is for GW190602$\_$175927: its remnant is retained with 60\% probability if the cluster has an escape speed of $600$ km/s. 
GW190521, the most massive BBH detected to date, also has a significant retention probability. A dense star cluster with escape speed $\sim 700$ km/s will be able to retain GW190521 with 50\% probability, potentially allowing  for the future formation of an intermediate mass BBH. 

We note that our method complements that of \citet{Kimball:2020qyd}. While their approach determines the hierarchical merger history of the binary constituents by measuring the masses and spins, our method projects these systems into the future and quantifies their ability to further participate in such mergers.

\section{\label{sec:charactercluster}Characterizing clusters via their retention of BH merger remnants}

\begin{figure}[t]
    \centering
    \includegraphics[scale=0.46]{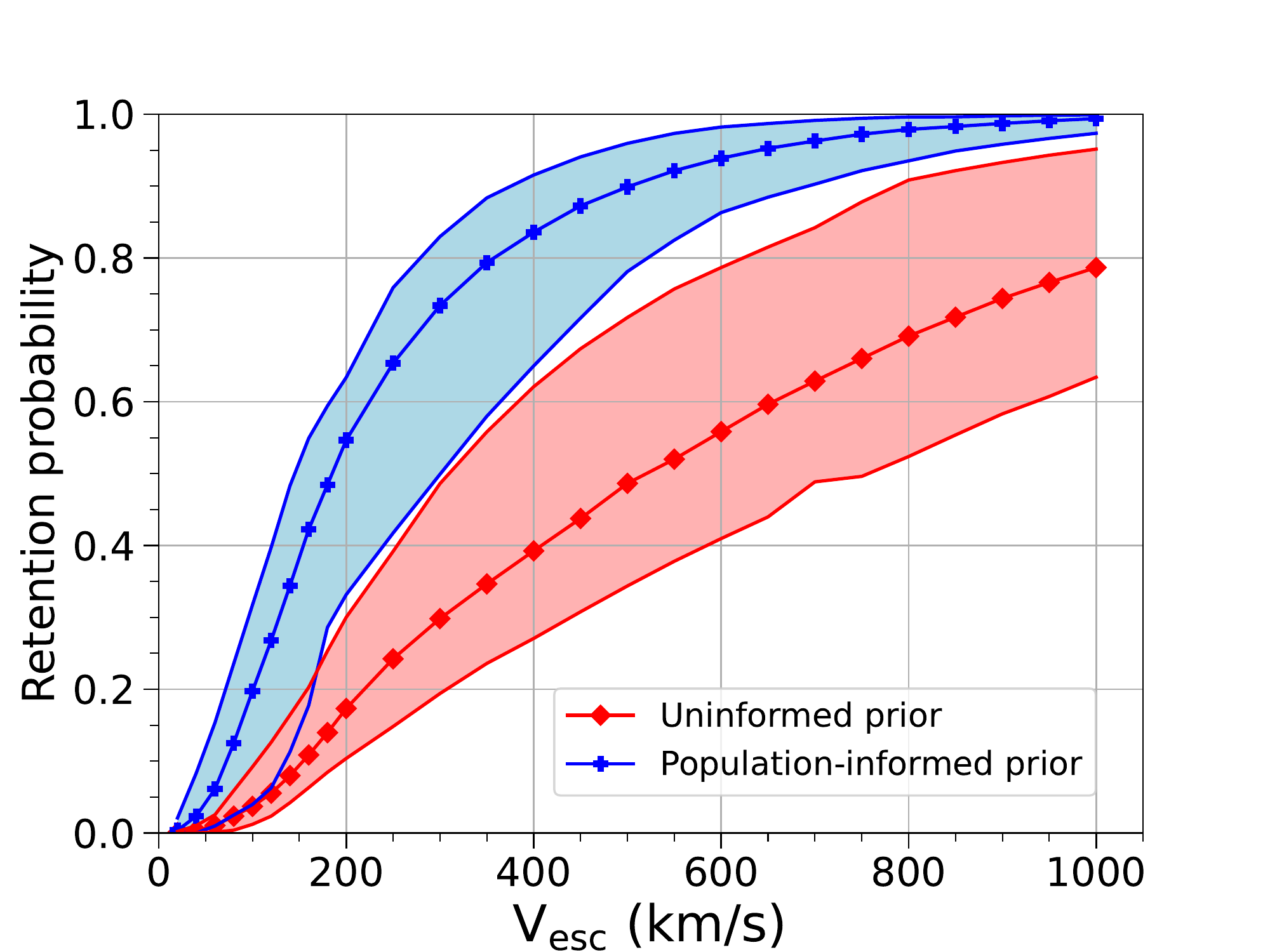}
    \caption{\label{fig:medianretention} {Distribution of retention probability, with and without population weighted samples, computed from all the informative events in GWTC-2 as a function of the escape speed of the star cluster $\rm{V_{esc}}$. The solid lines with markers show the median values while the shaded regions show the 90\% credible intervals of the distributions. These curves are obtained without any assumptions about the cluster type or properties (besides being characterized by a single value,  $\rm{V_{esc}}$).
    }
    } 
\end{figure}

 Having examined the retention probability of a given GWTC-2 event $j$, we now extend our results to the entire population of all informative GWTC-2 events. Let us suppose that all these informative events are characteristic of the population of BBHs that merge in a single star cluster that is solely parameterized by its escape speed $\rm{V_{esc}}$. Over time this entire population merges in  this cluster, and individual merger remnants are retained if ${\rm V}_{\rm kick}^j< {\rm V_{esc}}$. If the escape speed of the cluster ${\rm V_{esc}}$ is large compared to typical values of $\{{{\rm V}_{\rm kick}^j}\}$, a large fraction of the population is retained and can go on to seed the next generation of mergers. If not, then clusters with a particular value of ${\rm V_{esc}}$ (or lower) are likely to be poor sites for subsequent generations of BBH mergers. Here we will attempt to characterize if a particular cluster with escape speed ${\rm V_{esc}}$ is a good site for subsequent BBH mergers.

It is a strong assumption that the GWTC-2 BBHs form a sample reflective of the BBH population in a typical cluster parameterized only by its escape speed, and this assumption needs to be informed by future GW observations and cluster modeling. Besides, posteriors from the individual events do suffer from selection effects associated with gravitational wave observations~\citep{Fishbach:2019bbm,Moore:2021xhn}. Such selection effects may be mitigated on individual event posteriors with a careful reweighting via priors that are population-informed, as opposed to uniform, on different parameters~\citep{LIGOScientific:2020kqk}. This is achieved by invoking the population model inferred from the GWTC-2  and using the corresponding priors on various parameters for the analysis. The population-reweighted parameter estimation samples for different events are drawn from the \emph{GWTC-2  Data Release}~\citep{lvc:datadoi} corresponding to the Power-law + Peak mass model \citep{LIGOScientific:2020kqk}.

To characterize a particular cluster's suitability as a site of future hierarchical growth via its retention probability, we consider posteriors from individual events with uninformed priors and population-informed priors.
We look only into the 42 informative events based on the JS divergence criterion mentioned earlier, of which population-informed priors are available only for 40 of them (GW190719\_215514 and GW190909\_114149 are excluded). Hence, the results for population-informed priors are based on 40 events, whereas those for uninformative priors are based on 42 events.

 For a cluster with escape speed ${\rm V_{esc}}$, Section~\ref{sec:method} showed that a given event $j$ has a retention probability ${\rm P}_{{\rm ret},j}({\rm V_{esc}})$ (which, e.g., can be read off from Figure~\ref{fig:Retention0.5pc1}). If we consider that the entire set $\{j=1,\ldots,\, {\rm j_{max}}\}$ of informative GWTC-2 events is in a single cluster with ${\rm V_{esc}}$, and that the set is representative of the cluster's BBH population, then the samples for the distribution of retention probability for that cluster are given by the set $[{\rm P}_{{\rm ret},j}({\rm V_{esc}}),\; j=1,\ldots,\,{\rm j_{max}})]$, where ${\rm j_{max}}$ is $40$ or $42$ as discussed above. 
 This defines our distribution of retention probability, which is readily calculated from data like that shown in Figure~\ref{fig:Retention0.5pc1}.

 Using our calculated values for ${\rm P}_{{\rm ret},j}({\rm V_{esc}})$, we proceed to calculate the distribution of retention probability as a function of the cluster escape speed. In doing so, we make no assumption about the nature of the cluster. The result is shown in Figure~\ref{fig:medianretention}. The dots, which are joined, show the median retention probability and the shaded regions denote the 90\% credible interval.
 
With the uninformed priors, we find that clusters with escape speeds of 200 km/s and 500 km/s can retain $17^{+13}_{-7}$\% and $49^{+23}_{-14}$\% of the informative GWTC-2 population, while a cluster with an escape speed of about 800 km/s can retain $69^{+22}_{-17}$\% of the remnants.  When we use population-informed priors, we see that clusters with escape speeds of 200 km/s and 500 km/s can retain $55^{+9}_{-22}$\% and $90^{+6}_{-12}$\% of the informative GWTC-2 population, while a cluster with an escape speed of about 800 km/s can retain $98^{+2}_{-4}$\% of the remnants.
The numbers arising from the population-weighted analysis point to a significantly higher retention probability, thereby increasing the prospects of hierarchical mergers. Our estimates with population weighted samples broadly agree with those in \citet{Doctor:2021qfn}, which obtained retention fractions of $4\%$ and $40\%$ for clusters with escape speeds of 50 km/s  and 250 km/s.

An important implication of this result concerns the role of globular clusters, which have ${\rm V_{esc}} \lesssim 80 - 180$ km/s (with $180$ km/s at the high end of the escape speed distribution and most clusters under $80$ km/s; see Figure~3 of \citet{Antonini:2016gqe}).  These have median retention probabilities $\lesssim 14\%$ ($\lesssim 48\%$ with population weighted analysis). While a small subset of globular clusters could facilitate hierarchical mergers, the overall probability of globular clusters serving as major sites of hierarchical mergers is  low. However, nuclear star clusters, which have escape speeds $\gtrsim$200 km/s, can readily host hierarchical mergers as they retain more than half of the merger remnants.

\section{\label{sec:discussion}Retention probabilities of GWTC-2 events by globular and nuclear star clusters}
We now map our ``agnostic cluster'' curves in Figure~\ref{fig:medianretention} to globular and nuclear clusters, using their expected escape speed distributions~\citep{Harris:1996,Antonini:2016gqe,Georgiev:2016} to determine the probability that a particular GW event can be retained by a GC or NSC. (Note that these estimates of escape speeds are from the center of the cluster for GCs, whereas for NSCs these are defined at the half-mass radius of the cluster.)

We first construct a cluster escape speed PDF $p_k(\rm V_{esc})$, where the index $k$ denotes GC or NSC. This PDF is constructed using log-normal fits to the data in Figure~3 of \citet{Antonini:2016gqe}, which have means $\left\langle \log_{10}\left(\frac{{\rm V_{esc}}}{\rm km/s}\right) \right\rangle = (1.5, 2.2)$ and standard deviations $\sigma_{\log_{10}[{\rm V_{esc}}/({\rm km/s})]} = (0.3, 0.36)$ for GCs and NSCs (respectively).  The corresponding CDF (giving the probability that ${\rm V_{esc}} \leq v_{\ast}$) is denoted by $F_k(v_{\ast})$. Then the probability for a BH with kick $\rm V_{kick}$ to be retained by a given cluster is $1-F_k({\rm V_{kick}})$.

Given our kick PDF $p_j({\rm V_{kick}})$ for GWTC-2 event $j$, the probability that event $j$ is retained by a cluster of type $k$ is then
\begin{equation}
{\rm P}_{{\rm ret}, j}^{k} = \int_0^{{\rm V}_{\rm max}} p_j({\rm V_{kick}}) [1-F_k({\rm V_{kick}})] \, d{\rm V_{kick}}\,.
\label{eq:Pret}
\end{equation}
This simply amounts to weighting each bin of the kick PDF by the probability $[1-F_k({\rm V_{kick}})]$ that a cluster $k$ retains a kicked BH with ${\rm V_{kick}}$. 
We set ${\rm V}_{\rm max}=5000 \,{\rm km/s}$ (any choice ${\rm V}_{\rm max}\gtrsim 1000 \,{\rm km/s}$ would make only a negligible difference). It should be noted that Equation (\ref{eq:Pret}) assumes a uniform merger rate across the different clusters. In reality massive clusters, which have larger escape speeds, may have a higher merger rate. This may be folded into this estimate in the future.

Applying Equation~(\ref{eq:Pret}), the retention probability of GWTC-2 event $j$ by a GC or NSC is computed and given in Table~\ref{tab:Ret-GC-NSC}. These estimates are also given for both uninformed and population-informed priors. The highest retention probability is for GW190814 which is $\approx 82\%$ ($\approx 84\%$)  for uninformed 
(population-informed) priors for NSCs. This is due to the precise estimation of the low kick magnitude for this event. With uninformed priors, five events  have retention probabilities higher than 25\% for NSCs, whereas with population-informed samples 15 events  have retention probabilities higher than 45\% for NSCs. Similarly, the probability that GW190814 is retained by a GC, assuming the uninformed (population-informed prior) is $\approx 11\%$ ($\approx 13\%$). With the uninformed priors, four events have retention probabilities in GCs higher than 4\%, while the population-informed analysis yields 17 events with retention probabilities in GCs higher than 6\%.


\begin{figure}[t]
    \centering
   \includegraphics[scale=0.48]{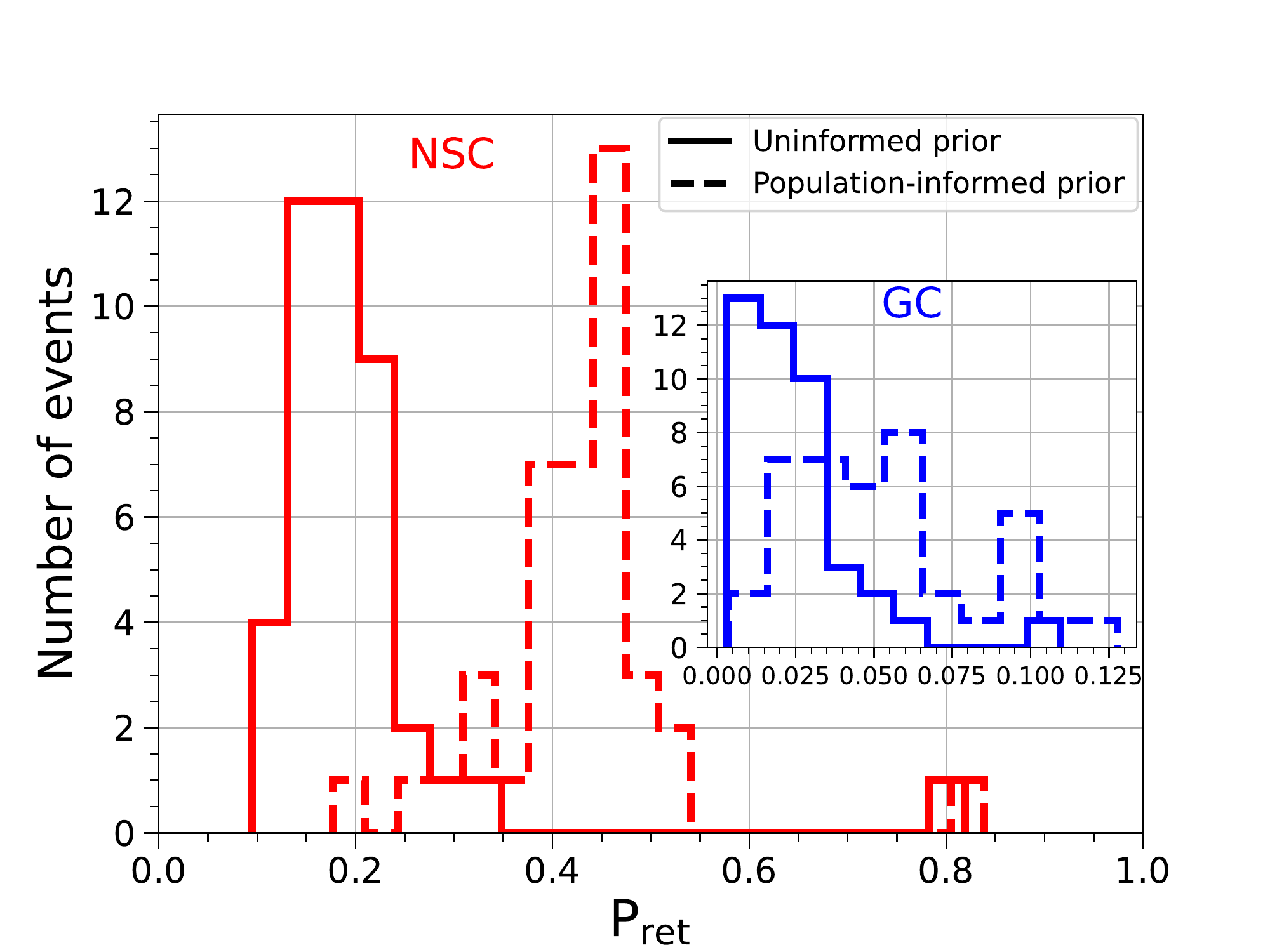}
    \caption{ \label{fig:Pret-GC-NSC}Distribution of retention probabilities for nuclear star clusters (red curves) and globular clusters (blue curves, inset) obtained from the informative events of GWTC-2. The distribution is obtained by binning the data in Table \ref{tab:Ret-GC-NSC}.
    With uninformed priors (solid curves) 90\% of the events have retention probability between 0.121-0.287 (0.006-0.056) for nuclear (globular) clusters. With population-informed priors (dashed curves), 90\% of the events have a retention probability between 0.304-0.514 (0.016-0.102) for nuclear (globular) clusters.
    }
\end{figure}

\begin{table*}[t]
 \caption{\label{tab:Ret-GC-NSC}{Probability of retention for all informative GWTC-2 events by GCs or NSCs using uninformed and population-informed (inside the parentheses) priors; see the text for details. This table uses Equation~(\ref{eq:Pret}) to compute the probability that a BH merger remnant is retained by a cluster, given the distribution of cluster escape speeds and the inferred kick posterior distribution for each GWTC-2 event.}
 }
\begin{center}
\begin{tabular}{|c|c|c|c|c|c|}
\hline
\multirow{3}{*}{Event name} & \multicolumn{2}{c|}{Retention probability} & \multirow{3}{*}{Event name} & \multicolumn{2}{c|}{Retention probability}\\
\cline{2-3}\cline{5-6}
 & GC & NSC & & GC & NSC\\
\hline
    GW150914 &  0.0563 (0.0967) &  0.2880 (0.4780) & GW190602$\_$175927 &  0.0097 (0.0455) &  0.1795 (0.4139)\\
    GW151226 &  0.0216 (0.0273) &  0.1645 (0.3069) & GW190620$\_$030421 &  0.0110 (0.0237) &  0.1276 (0.3150)\\
    GW170104 &  0.0331 (0.0313) &  0.2192 (0.4190) & GW190630$\_$185205 &  0.0357 (0.0320) &  0.2385 (0.4004)\\
    GW170729 &  0.0182 (0.0335) &  0.1413 (0.3298) & GW190701$\_$203306 &  0.0161 (0.0749) &  0.1875 (0.4745)\\
    GW170814 &  0.0345 (0.0440) &  0.1980 (0.4429) & GW190706$\_$222641 &  0.0124 (0.0226) &  0.1603 (0.3712)\\
    GW170818 &  0.0271 (0.0373) &  0.1724 (0.3936) & GW190708$\_$232457 &  0.0489 (0.1058) &  0.2728 (0.5189)\\
    GW170823 &  0.0291 (0.0638) &  0.1897 (0.4570) & GW190719$\_$215514 &  0.0136 &  0.1432 \\
    GW190408$\_$181802 &  0.0383 (0.0655) &  0.2247 (0.4700) & GW190720$\_$000836 &  0.0250 (0.0413767) &  0.2125 (0.4069)\\
    GW190412 &  0.0031 (0.0037) &  0.1903 (0.2442) & GW190727$\_$060333 &  0.0238 (0.0916) &  0.1768 (0.4929)\\
    GW190413$\_$052954 &  0.0274 (0.0940) &  0.1939 (0.4540) & GW190728$\_$064510 &  0.0311 (0.0478) &  0.2278 (0.4444)\\
    GW190413$\_$134308 &  0.0202 (0.0464) &  0.1658 (0.4104) & GW190731$\_$140936 &  0.0252 (0.0652) &  0.1757 (0.4567)\\
    GW190421$\_$213856 &  0.0130 (0.0770) &  0.1662 (0.4706) & GW190803$\_$022701 &  0.0145 (0.0612) &  0.1961 (0.4463)\\
    GW190424$\_$180648 &  0.0192 (0.0585) &  0.1534 (0.4371) & GW190814 &  0.1097 (0.1276) &  0.8190 (0.8384)\\
    GW190503$\_$185404 &  0.0112 (0.0642) &  0.1930 (0.4757) & GW190828$\_$063405 &  0.0181 (0.0322) &  0.1614 (0.3811)\\
    GW190512$\_$180714 &  0.0344 (0.0261) &  0.2735 (0.4255) & GW190828$\_$065509 &  0.0204 (0.0216) &  0.2226 (0.3905)\\
    GW190514$\_$065416 &  0.0117 (0.0656) &  0.1647 (0.4674) & GW190909$\_$114149 &  0.0098 &  0.1658\\
    GW190517$\_$055101 &  0.0092 (0.0168) &  0.0947 (0.1768) & GW190910$\_$112807 &  0.0361 (0.0992) & 0.2217 (0.5133)\\
    GW190519$\_$153544 &  0.0060 (0.0090) &  0.1096 (0.3138) & GW190915$\_$235702 &  0.0138 (0.0790) &  0.1519 (0.4534)\\
    GW190521 &  0.0071 (0.1023) &  0.1201 (0.4732) & GW190924$\_$021846 &  0.0583 (0.0387) &  0.3216 (0.4456)\\
    GW190521$\_$074359 &  0.0174 (0.0620) &  0.1854 (0.4278) & GW190929$\_$012149 &  0.0049 (0.0241) &  0.2080 (0.3928)\\
    GW190527$\_$092055 &  0.0142 (0.0399) &  0.1615 (0.4041) & GW190930$\_$133541 &  0.0299 (0.0506) &  0.2130 (0.4208)\\
\hline
\end{tabular}
\end{center}
\end{table*}


Using the numbers in Table~\ref{tab:Ret-GC-NSC}, we construct in  Figure~\ref{fig:Pret-GC-NSC} the distribution of retention probabilities for NSCs and GCs. If all the informative GWTC-2 mergers happened in NSCs, we find that the retention probabilities of 90\% of those events would be between $0.121$ and $0.287$ ($0.304$--$0.514$) based on uninformed (population-informed) priors. Similarly, if all informative GWTC-2 events are associated with globular clusters, 90\% of them will have retention probabilities between $0.006$ and $0.056$ ($0.016$--$0.102$) using uninformed (population-informed) priors.
The outlier bump at large ${\rm P_{ret}}$ in Figure~\ref{fig:Pret-GC-NSC} is due to GW190814; with the smallest (and best-measured) kick magnitude, it yields the largest retention probability.

The area of the two distributions in Figure~\ref{fig:Pret-GC-NSC} (or summing the values in Table~\ref{tab:Ret-GC-NSC}) gives us an estimate of the number of remnants that will be retained out of the 42 (40) events, corresponding to the uninformed (population-informed) priors, we considered in the GWTC-2 population. There are $\sim 9$ ($\sim 1$) that could participate in further mergers if  those events happened in NSCs (GCs), using uninformed priors. The same numbers for the population-informed priors are   $\sim17$ ($\sim2$). The estimates based on standard single event analysis are in good agreement with the projections of \cite{Mapelli:2020xeq}, which indicate that roughly 15\% (0.6\%) of mergers in NSCs (GCs) would be hierarchical in nature. Note that these are present-day ($z=0$) retention probabilities; given the limited redshift reach of GWTC-2, we have not considered the redshift evolution of retention probability.

\section{Conclusion} 
 Formation and growth of black holes at different scales is an important open problem in astrophysics. Hierarchical mergers are one of the channels considered for black hole growth, although its efficiency is largely unknown. Observations of binary black hole mergers by LIGO/Virgo 
 provide one of the most important tools to probe the hierarchical growth of black holes at the stellar mass scale.  Predicted mass and spin distributions from hierarchical mergers were compared with GWTC-2 events in \cite{Kimball:2020qyd} to assess the presence of hierarchical mergers. We discussed another approach for evaluating the prospects of hierarchical growth, by computing the probability that GWTC-2 merger remnants are retained by cluster environments---a necessary condition for hierarchical growth. This involved computing the gravitational-wave kick probability distribution for each GWTC-2 event and comparing with cluster escape speeds. Using population-informed posterior samples, we found that among the GWTC-2 merger remnants,  17 remnants are expected to be retained by their host-clusters if they all merged in a nuclear star cluster, or about two events if all GWTC-2 mergers occurred in globular clusters. These retained remnants could potentially participate in a subsequent generation of binary black hole mergers, provided favorable circumstances for binary formation exist near these black hole remnants. As hierarchical mergers are assisted by larger cluster escape speeds and a higher frequency of in-cluster mergers, understanding the redshift evolution of these two quantities will be important in determining the efficiency of hierarchical mergers as a function of cosmic time.

Our conclusions assume that all GWTC-2 mergers happened in star clusters and not in galactic fields, and this is hard to verify at present. 
As enhanced detector sensitivities allow the discovery of more binary black holes with improved parameter accuracy and precision, the methods discussed here will provide a unique probe of the hierarchical growth of stellar-mass black holes.

\acknowledgments
We thank Fabio Antonini for reading the manuscript and providing useful comments. We also thank J.~Bustillo and V.~Varma for comments on the paper, S.~Chatterjee for valuable discussions and pointers to important references, and S.~Datta, P.~Saini, P.~Ajith, S.~Kapadia, A.~Vijayakumar and M.~A.~Shaikh for useful discussions. K.G.A. acknowledges the Swarnajayanti grant DST/SJF/PSA-01/2017-18 of the Department of Science and Technology, India. K.G.A and P.M. acknowledge the support of the Core Research Grant EMR/2016/005594 of the Science and Engineering Research Board of India and a grant from the Infosys foundation.  
This material is based upon work supported by the NSF's LIGO Laboratory which is a major facility fully funded by the National Science Foundation. We also acknowledge NSF support via NSF CAREER award PHY-1653374 to M.F. and NSF awards AST-2006384, PHY-2012083 and PHY-1836779 to B.S.S. This research has made use of data obtained from the Gravitational Wave Open Science Center (www.gw-openscience.org), a service of LIGO Laboratory, the LIGO Scientific Collaboration and the Virgo Collaboration. 
Virgo is funded by the French Centre National de Recherche Scientifique (CNRS), the Italian Istituto Nazionale della Fisica Nucleare (INFN), and the Dutch Nikhef, with contributions by Polish and Hungarian institutes.
This has the LIGO preprint number P2100188.

\clearpage
\appendix
\section{\label{sec:kick-appen}Kick fitting formula} 
Here we summarize the employed model for inferring the magnitude of the kick velocity. We directly apply fitting formulas based on numerical relativity simulations of binary black holes~\citep{Gonzalez:2006, Campanelli:2007ew, Gonzalez:2007hi, Lousto:2007, Lousto:2012su, Lousto:2013}. Following \citet{Campanelli:2007ew}, the kick velocity magnitude is decomposed as
\begin{equation}
\label{eq:kick}
{\rm V}_{\rm kick} = \sqrt{ V_{m}^{2} + 2 V_{m} V_{s \perp} \cos\xi + V_{s \perp}^{2} + V_{s \parallel}^{2}} \hspace{0.05 cm},
\end{equation}
where $V_{m}$, $V_{s \perp}$, $V_{s \parallel}$ are expressed in terms of
\begin{eqnarray}
\boldsymbol\Delta = \frac{\boldsymbol{\chi_{1}} - q \boldsymbol{\chi_{2}} }{1+q},\\ \label{eq:delta}
\widetilde{\boldsymbol\chi} =  \frac{q^{2} \boldsymbol{\chi_{2}}  + \boldsymbol{\chi_{1}}}{(1+q)^{2}},\label{eq:chi-tilde}
\end{eqnarray}
where $q \le 1$ is the mass ratio of the binary and $\boldsymbol{\chi_{1,2}}$ are the dimensionless spin vectors of the progenitor BHs \citep{Lousto:2012su, Lousto:2013, Gerosa:2016sys}. Contributions to the kick velocity listed in Equation~(\ref{eq:kick}) are given by~\citep{Lousto:2012su, Lousto:2013}: 
\begin{eqnarray}
 V_{m}&=& A \,\eta^{2} \frac{1-q}{1+q} (1 + B \,\eta ),\\ \label{eq:kick-mass}
 V_{s \perp}&=& H \, \eta^{2} \, \Delta_{\parallel},\\ \label{eq:kick-perp}
 V_{s \parallel}&=& 16 \,\eta^{2} [\Delta_{\perp}(V_{11} + 2 V_{A} \widetilde{\chi}_{\parallel} + 4 V_{B} \widetilde{\chi}_{\parallel}^{2} \nonumber\\ 
 &+& 8 V_{C} \widetilde{\chi}_{\parallel}^{3}) + 2 \widetilde{\chi}_{\perp} \Delta_{\parallel} (C_{2} + 2 C_{3} \widetilde{\chi}_{\parallel})] \cos\Theta,\label{eq:kick-parallel} 
\end{eqnarray}
where 
\begin{eqnarray}
 \eta&=&\frac{q}{(1+q)^{2}},\\
 \Delta_{\parallel}&=& |\boldsymbol \Delta \cdot \hat{\textbf{L}}|=\frac{\chi_{1} \cos \theta_{1} - q \chi_{2} \cos \theta_{2}}{1+q},\\
 \Delta_{\perp}&=& |\boldsymbol\Delta \times \hat{\textbf{L}}|\nonumber\\ 
 &=&\frac{1}{1+q} [\chi_{1}^{2} \sin^{2}\,\theta_{1} + q^{2}\, {\chi_{2}}^{2} \sin^{2} \theta_{2}\nonumber\\
 &&-2 q\chi_{1}\,\chi_{2} \sin \theta_{1} \sin \theta_{2} \cos\phi_{12}]^{1/2}\,,\\
 \widetilde{\chi}_{\parallel}&=& \widetilde{\boldsymbol\chi} \cdot \hat{\textbf{L}}=\frac{\chi_{1} \cos \theta_{1} + q^{2} \chi_{2} \cos \theta_{2}}{(1+q)^{2}} \, , \; \; \text{and} \\
 \widetilde{\chi}_{\perp}&=& |{\boldsymbol\chi} \times \hat{\textbf{L}}| \nonumber\\
 &=& \frac{1}{(1+q)^2} [{\chi_{1}}^{2} \sin^{2} \theta_{1} + q^{4} {\chi_{2}}^{2} \sin^{2} \theta_{2}  \nonumber\\
 &&+ 2 q^{2} \chi_{1} \chi_{2} \sin \theta_{1} \sin \theta_{2} \cos \phi_{12}]^{1/2}\,.
\end{eqnarray}
In the above $\hat{\textbf{L}}$ is the unit vector along the total orbital angular momentum, $\theta_{1,2}$ are the angles between $\hat{\textbf{L}}$ and $\boldsymbol{{\chi}_{1,2}}$, $\phi_{12}$ is the difference between the azimuthal angles of the two spin vectors, and $\chi_{1,2}$ are the spin magnitudes. The values of the different numerical fitting coefficients are $A=1.2 \times 10^{4}$ km/s, $B=-0.93$ \citep{Gonzalez:2006}, $H=6.9 \times 10^{3}$ km/s \citep{Lousto:2007}, $V_{11}=3677.76$ km/s, $V_{A}=2481.21$ km/s, $V_{B}=1792.45$ km/s, $V_{C}=1506.52$ km/s \citep{Lousto:2012}, $C_{2} = 1140$ km/s, and $C_{3} = 2481$ km/s \citep{Lousto:2013}. The angle $\xi \sim 145^{\circ}$ for a wide range of quasi-circular configurations \citep{Lousto:2007}, whereas $\Theta$ is defined as the angle between ${\bm \Delta} \times \hat{{\bm L}}$ and a fiducial infall direction of the two holes at merger. (See \citet{Gerosa:2016sys} for a more precise definition of $\Theta$, which depends on the orbital phase of the binary at an arbitrary reference time.)

Since the spin vectors can evolve considerably throughout the binary inspiral, the kick velocity fitting formula can only be employed close to the merger, at separations $\sim 10 M$ where NR simulations typically start (see \cite{Barausse:2009uz} and Section 2.1 of \cite{Gerosa:2014gja}). To account for this, we evolve the posterior samples (which are specified at a reference frequency of $20\, {\rm Hz}$) to a separation of $10M$ using orbit-averaged post-Newtonian equations of motion \citep{ACST94,K95,Racine:2008qv} contained in the LALSimulation libraries of LALSuite \citep{lalsuite}. (For GW190521, the reference frequency is chosen to be $11$ Hz.)

For BBHs from the first two observing runs (GWTC-1),  we use the posterior samples produced by the Bayesian inference package {\tt Bilby}~\citep{Ashton:2018jfp,Romero-Shaw:2020owr} using the {\tt IMRPhenomPv2} waveform model~\citep{Hannam:2013oca}. For events observed during the first half of the third observing run (reported in GWTC-2), posterior samples are drawn from the  \emph{GWTC-2 Data Release}~\citep{lvc:datadoi} corresponding to the {\tt SEOBNRv4PHM} waveform model \citep{Cotesta:2018fcv}; {\tt SEOBNRv4P} \citep{Ossokine:2020kjp} samples are used for those events for which {\tt SEOBNRv4PHM} samples are not available.

\section{\label{sec:JSdiv}Jensen-Shannon divergence for the kick posteriors}
The issue of prior-dominated posteriors in the context of kick magnitude inference was pointed out by \cite{Varma:2020nbm} while inferring the kicks for GW150914 and GW170729. Priors on the kick magnitude are constructed by combining priors on the mass ratio and spin parameters with relevant NR fitting formulas for the kick. It was specifically noted that the difference between the inferred kick posteriors and the corresponding priors is very small, as evidenced by the Kullback–Leibler (KL) divergence~\citep{KLdiv} between them. This was attributed to the lack of stringent constraints on the spin parameters of the binary. 
Here, in order to assess the information content of our kick posteriors relative to our priors, we employ a similar measure, the Jensen–Shannon (JS) divergence~\citep{JSdiv}. This is a symmetric (with respect to the two distributions that are being compared) version of the KL divergence and always has a finite value. We set a threshold of $0.007$ for the JS divergence (the same threshold used in \citet{GWTC2} for comparing posterior distributions between different waveform approximants); above this threshold we consider the posteriors to be informative. For technical details of JS divergence, see Appendix A of \cite{GWTC2}.

This criterion reduces our GWTC-2 sample to 42 informative binaries. We restrict this study to these informative events. Note that we do not quote the JS divergence for the population-informed analysis as the 40 events we analyze under this assumption will, by definition, cross this threshold as they carry more information. Table~\ref{tab:JS-div} lists the JS divergence for the events that crossed the $0.007$ threshold. GW190814 has the highest JS divergence and also has the most informative kick posterior in the entire GWTC-2 catalog. 

\begin{table*}[t]
  \begin{center}
    \caption{JS divergence between the posterior and prior distributions of the kick magnitude for different events in GWTC-2 using uninformed priors. Higher values of JS divergence imply more informative kick posteriors, with GW190814 being the most informative one in GWTC-2. Following \cite{GWTC2}, a threshold of 0.007 on the JS divergence is imposed, and only those events that pass the threshold are presented in the table. For a Gaussian distribution, this threshold corresponds to a 20\% shift in its mean.}
    \begin{tabular}{|l|c|c|c|c|c|c|l|}
    \hline
    Event name & JS divergence & Event name & JS divergence & Event name & JS divergence\\ 
    \hline\hline
     GW190814  & 0.4329             & GW190521$\_$074359 & 0.0460  & GW170729 & 0.0203\\
     GW190412 & 0.1579             & GW151226 & 0.0408   & GW190512$\_$180714 & 0.0190\\
     GW190924$\_$021846 & 0.1158   & GW190620$\_$030421 & 0.0405   & GW190408$\_$181802 & 0.0159\\
     GW190521  & 0.1092    & GW190602$\_$175927 & 0.0401  & GW190708$\_$232457 & 0.0159\\
     GW190517$\_$055101 & 0.0917   & GW190720$\_$000836 & 0.0364   & GW190727$\_$060333 & 0.0153\\
     GW190514$\_$065416 & 0.0913   & GW190929$\_$012149 & 0.0361  & GW190527$\_$092055 & 0.0133\\
     GW190930$\_$133541 & 0.0900            & GW170818 & 0.0301  & GW190701$\_$203306 & 0.0114\\
     GW190909$\_$114149 & 0.0605   & GW190424$\_$180648 & 0.0300  & GW190910$\_$112807 & 0.0114\\
     GW190728$\_$064510 & 0.0588             & GW190731$\_$140936 & 0.0293  & GW190503$\_$185404 & 0.0108\\
     GW190828$\_$065509 & 0.0560          & GW190421$\_$213856 & 0.0288            & GW190706$\_$222641 & 0.0100\\
     GW190413$\_$134308 & 0.0521  & GW190828$\_$063405 & 0.0281  &  GW150914 & 0.0090\\
     GW190719$\_$215514 & 0.0471  & GW190803$\_$022701 & 0.0261   & GW170104 & 0.0086\\
     GW190519$\_$153544 & 0.0471   & GW190630$\_$185205 & 0.0242  & GW190413$\_$052954 & 0.0084\\
     GW190915$\_$235702 & 0.0467  & GW170814 & 0.0224  & GW170823 & 0.0082\\
     \hline\hline
    \end{tabular}
    \label{tab:JS-div}
  \end{center}
\end{table*}

\bibliographystyle{aasjournal}
\bibliography{ref-list}
\end{document}